\documentclass[12pt]{article}
\usepackage{amsmath}
\usepackage{bbm,amssymb}

\newcommand{\eq}[1]{\begin{equation}#1\end{equation}}
\newcommand{\spl}[1]{\begin{split}#1\end{split}}
\newcommand{\al}[1]{\begin{align}#1\end{align}}

\usepackage{amsmath,latexsym}
\usepackage{graphicx}
\usepackage{psfrag}
\usepackage[latin1]{inputenc}
\usepackage{subfigure}
\usepackage{nicefrac}
\usepackage[stable]{footmisc}
\usepackage{comment}

\newcommand\nn{\nonumber}
\newcommand{\bg}[1]{\hat{#1}}

\def\d{\text{d}}

\def\slashchar#1{\setbox0=\hbox{$#1$}           % set a box for #1
\dimen0=\wd0                                 % and get its size
\setbox1=\hbox{/} \dimen1=\wd1               % get siste of /
\ifdim\dimen0>\dimen1                        % #1 is bigger
\rlap{\hbox to \dimen0{\hfil/\hfil}}      % so center / in box
#1                                        % and print #1
\else                                        % / is bigger
\rlap{\hbox to \dimen1{\hfil$#1$\hfil}}   % so center #1
/                                         % and print /
\fi}

\def\Re           {{\rm Re\hskip0.1em}}
\def\Im           {{\rm Im\hskip0.1em}}

\newcommand{\E}{\text{\tiny E}}

\def\Re           {{\rm Re\hskip0.1em}}
\def\Im           {{\rm Im\hskip0.1em}}

\begin{document}
\font\cmss=cmss10 \font\cmsss=cmss10 at 7pt

\vskip -0.6cm
\rightline{\small{\tt MPP-2008-139}}

\vskip .7 cm
%\hfill IC/2004/ \vskip .1in \hfill CPHT \vskip .1in \hfill hep-th/yymmnnn

\hfill
\vspace{18pt}
\begin{center}
{\Large \textbf{On the effective theory of type IIA AdS$_4$ compactifications}}
\end{center}

\vspace{6pt}
\begin{center}
{\large\textsl{Simon K\"ors}}

\vspace{25pt}
\textit{\small Max-Planck-Institut f\"{u}r Physik -- Theorie,\\
                    F\"{o}hringer Ring 6,  D-80805 M\"{u}nchen, Germany}
  \vspace{6pt}
\end{center}

\vspace{12pt}

\begin{center}
\textbf{Abstract}
\end{center}

\vspace{4pt} This is a summary of \cite{adseff}, where the low energy effective theory of type IIA AdS$_4$ $\mathcal{N}=1$ flux compactifications on nilmanifolds and cosets has been analyzed. We compute the superpotential, the K\"ahler potential and the mass spectrum for the light moduli. For the nilmanifold examples we perform a cross-check on the result for the mass spectrum by calculating it from a direct Kaluza-Klein reduction.

\vspace{1cm}

\noindent {\em Contribution to the proceedings of the 4th workshop of the RTN project `Constituents, Fundamental Forces and Symmetries of the Universe' in Varna, 11-17 September, 2008.}

%\thispagestyle{empty}

%\vfill
%\vskip 5.mm
%\hrule width 5.cm
%\vskip 2.mm
%{\small
%\noindent e-mail: koers@mppmu.mpg.de
%}

%\newpage
\setcounter{footnote}{0}

\section{Supersymmetric type IIA AdS$_4$ compactifications}

To date all our explicit ten-dimensional examples of ${\cal N}=1$ supersymmetric compactifications to AdS$_4$ fall within the class of type IIA SU(3)-structure compactifications and T-duals thereof. In many of these examples one \emph{needs} supersymmetric sources in order to satisfy the tadpole conditions but in all of them one \emph{can} add them. The general properties of supersymmetric sources and their consequences for the integrability of the supersymmetry equations were discussed in \cite{kt} within the framework of generalized geometry. Extending the work of \cite{lt}, it was shown in this reference that (under certain mild assumptions) all the equations of motion, appropriately source-modified, are automatically satisfied, if the supersymmetry
conditions in the bulk as well as for the sources together with the source-modified Bianchi identities are satisfied. This means one only has to solve the supersymmetry conditions in the bulk and the source-modified Bianchi identities for a supersymmetric (generalized calibrated \cite{gencal1,gencal2}) source in order to find a supersymmetric 'vacuum' , i.e. a particular solution of the equations of motion of the ten-dimensional supergravity.

\subsection{Conditions for a supersymmetric vacuum}
\label{IIAsusycond}

We assume the following $\mathcal{N}=1$ compactification ansatz for the ten-dimensional supersymmetry generators \cite{granaN1}:
\eq{\label{spinansatz}\spl{
\epsilon_1 & =\zeta_+\otimes \eta^{(1)}_+ \, + \, \zeta_-\otimes \eta^{(1)}_- \ , \qquad
\epsilon_2  =\zeta_+\otimes \eta^{(2)}_\mp \, + \, \zeta_-\otimes \eta^{(2)}_\pm \ ,
}}
with upper/lower sign for IIA/IIB, where $\zeta_\pm$ are four-dimensional and $\eta^{(1,2)}_\pm$ six-dimensional Weyl spinors. The Majorana conditions for $\epsilon_{1,2}$ imply the four- and six-dimensional reality conditions $(\zeta_+)^*=\zeta_-$ and $(\eta^{(1,2)}_+)^*=\eta^{(1,2)}_-$. This reduces the structure of the {\em generalized} tangent bundle to SU(3)$\times$SU(3) \cite{gualtieri}. The supersymmetry generators $\eta^{(1)}$ and $\eta^{(2)}$ can be combined into two spinor bilinears which (using the
Clifford map) can be associated with two polyforms of definite degree
\eq{
\slashchar{\Psi}_+ = \frac{8}{|a||b|}\eta^{(1)}_+ \otimes \eta^{(2)\dagger}_+ \, , \qquad \qquad
\slashchar{\Psi}_- = \frac{8}{|a||b|}\eta^{(1)}_+ \otimes \eta^{(2)\dagger}_- \, .
}
Redefining
$\Psi_1 = \Psi_{\mp} \,$, $\Psi_2 = \Psi_{\pm}$,
the supersymmetry conditions (in string frame) take the following concise form \cite{granaN1} (as usual $\Phi$ is the dilaton and $e^{A}$ the warp factor):
\begin{eqnarray}
\label{gensusy}
 \d_H \left( e^{4A-\Phi} \Im \Psi_1 \right) &=& 3 e^{3A-\Phi} \Im (W^* \Psi_2) + e^{4A} \tilde{F} \, , \nn \\
 \d_H \left[ e^{3A-\Phi} \Re (W^* \Psi_2) \right] &=& 2 |W|^2 e^{2A-\Phi} \Re \Psi_1 \, , \nn \\
 \d_H \left[ e^{3A-\Phi} \Im (W^* \Psi_2) \right] &=& 0 \, ,
\end{eqnarray}
where $F \equiv \hat{F} + \text{vol}_4 \wedge \tilde{F}$ 
and $W$ is defined in terms of the AdS Killing spinors: $\nabla_\mu \zeta_- = \pm \frac{1}{2} W \gamma_\mu \zeta_+$ for IIA/IIB.
{} From the above, the equations of motion for $F$ follow as integrability conditions.

When the internal supersymmetry generators of \eqref{spinansatz} are proportional,
$
\eta^{(2)}_+ = (b/a) \eta^{(1)}_+ 
$,
with $|\eta^{(1)}|^2 = |a|^2, |\eta^{(2)}|^2=|b|^2$,
the structure group reduces to SU(3). To compute the polyforms we define a normalized
spinor $\eta_+$ such that $\eta^{(1)}_+= a \eta_+$ and $\eta^{(2)}_+ = b \eta_+$
and moreover we choose the phase of $\eta$ such that $a=b^*$. It turns out that in compactifications
to AdS$_4$ the supersymmetry imposes $|a|^2=|b|^2$ such that $b/a=e^{i\theta}$ is just a phase.
For the polyforms we get
\eq{
\label{SU3pure}
\Psi_+ = - \Omega \, , \quad \Psi_- = e^{-i\theta} e^{iJ}  \, 
\qquad \text{with} \quad
J_{mn}\equiv i \eta^{\dagger}_+ \gamma_{mn} \eta_+ \, , \quad \Omega_{mnp}\equiv \eta^{\dagger}_- \gamma_{mnp} \eta_+ \, .
}
The real non-degenerate two-form $J$ and the complex decomposable
three-form $\Omega$ completely specify an SU(3)-structure on the six-dimensional
 manifold $\mathcal{M}$ because they satisfy $\Omega\wedge J=0$ and 
$\Omega\wedge\Omega^*=\frac{4i}{3}J^3\neq 0$, and the associated metric is positive definite. The intrinsic torsion of a manifold with $SU(3)$-structure decomposes into five torsion classes which also appear in the
SU(3) decomposition of the exterior derivative of $J$ and $\Omega$:
\eq{\spl{
\d J=\frac{3}{2}\Im(\mathcal{W}_1\Omega^*)+\mathcal{W}_4\wedge J+\mathcal{W}_3 \, , \qquad
\d \Omega= \mathcal{W}_1 J\wedge J+\mathcal{W}_2 \wedge J+\mathcal{W}_5^*\wedge \Omega ~.
\label{torsionclassesv}
}}
As we will show,  in the vacua of interest to us only the classes $\mathcal{W}_1$, $\mathcal{W}_2$ are non-vanishing and they are
purely imaginary, which we will indicate with a minus superscript: $\mathcal{W}_{1,2}=\mathcal{W}^-_{1,2}=i\Im\mathcal{W}^-_{1,2}$.

Plugging \eqref{SU3pure} and \eqref{torsionclassesv} into \eqref{gensusy} one gets
the most general form of $\mathcal{N}=1$ compactifications of IIA supergravity to AdS$_4$ with SU(3)-structure, which was originally derived in \cite{lt}. The dilaton and the warp factor have to be constant and setting the latter to one, the solutions are given by:
%%%%%
\begin{eqnarray}
\label{ltsol}
F_2 &=&\frac{f}{9}J+F'_2 \, , \qquad \qquad \qquad
F_4~~=~~f\mathrm{vol}_4+\frac{3m}{10} J\wedge J \, , \nn \\
H &=&\frac{2m}{5} e^{\Phi}\Re \Omega \, , \qquad \qquad 
W e^{i \theta} ~~=~~-\frac{1}{5} e^{\Phi}m+\frac{i}{3} e^{\Phi }f \, .
\end{eqnarray}
%%%%%
where $H$ is the NSNS three-form, and $F_{n}$ denote the RR forms. The constants $f$ and $m$  parameterize the solution: $f$ is the Freund-Rubin parameter, while $m$
is the mass of Romans' supergravity \cite{roma} -- which can be identified with $F_0$ in the `democratic'
formulation \cite{democratic}.
% the phase $\alpha$ parameterizes the `gauge-freedom'
%stemming from the fact that the SU(3)-structure is only defined up to U(1) transformations. In \cite{lt}
%$\alpha$ was gauge-fixed to zero; here we have chosen to restore it for later convenience.
The two-form $F'_2$ is the primitive part of $F_2$ (i.e.\ it is in the $\bf{8}$ of SU(3)). 
The only non-zero torsion classes of the internal manifold are ${\cal W}^-_1,{\cal W}^-_2$
and they are given by:
\eq{
{\cal W}^-_1=-\frac{4i}{9} e^{\Phi} f  \, , \qquad
{\cal W}^-_2=-i e^{\Phi} F'_2 \,  .
\label{ltsolc}
} 
The only condition from the Bianchi identities is
\eq{
\d F'_2=(\frac{2}{27}f^2-\frac{2}{5}m^2 ) e^{\Phi} \Re \Omega - j^{6} ~,\label{ltsolb}
}
where $j^6$ is a possible \emph{smeared} six-brane/plane, whose form is constrained by its calibration conditions:
\eq{\label{orinonprop}
j^6 \wedge \Re \Omega = 0 \, , \qquad j^6 \wedge J = 0 
 \qquad \Rightarrow \qquad
j^{6}=-\frac{2}{5}e^{-\Phi}\mu \Re\Omega + w_3~,
}
with $w_3$ a primitive (2,1)+(1,2)-form. 
If one plugs \eqref{ltsolc} and \eqref{orinonprop} into \eqref{ltsolb}, one gets
\begin{eqnarray}\label{bound}
w_3 = -i e^{-\Phi} \d \mathcal{W}_2^-\Big|_{(2,1)+(1,2)} \, , \qquad
\label{boundtra} e^{2\Phi} m^2=\mu+\frac{5}{16}\left(3|{\cal W}^-_1|^2-|{\cal W}^-_2|^2\right) \ge 0 \, . 
\end{eqnarray}
Instead of $m$ one can use $\mu$  as a parameter of the solution.

To summarize: In order to find $\mathcal{N}$ = 1 supersymmetric AdS$_4$ vacua of type IIA  supergravity  on manifolds with $SU(3)$-strucure, it suffices to verify that the torsion classes ${\cal W}_{3,4,5}$ are vanishing. The solution is then given by \eqref{ltsol}, where one has to use \eqref{ltsolc}.  The source can be read off from (\ref{orinonprop}) and \eqref{bound}.

\subsection{Hierarchy of scales}

To promote a given supergravity vacuum to a trustworthy approximation of a string theory vacuum, we need to show that we can consistently take the string coupling constant
to be small ($g_s=e^{\Phi}\ll 1$), so that string loops can be safely ignored, and that the volume of the internal manifold  is large in string units ($L_{int}/l\gg 1$, where $L_{int}$ is the characteristic length of the internal manifold), so that $\alpha'$-corrections can be neglected. 
Following \cite{tomtwistor} it is shown in \cite{adseff} that this is possible in all our models.

A further requirement is that we can decouple the Kaluza-Klein tower ($
|\Lambda_{\text{AdS}}|L_{int}^2 \ll 1$) in order to make the analysis of the effective theory in section \ref{lowennil} selfconsistent. Taking into account $|\Lambda_{\text{AdS}}|\sim |W|^2$ we find from \eqref{ltsol} the condition
\eq{
\label{Wsmall}
 \frac{1}{25} (g_s)^2 m^2 L_{int}^2 + \frac{1}{9} (g_s)^2 f^2 L_{int}^2 \ll 1 \, ,
}
which means that each of the two terms must be separately much
smaller than one. 
We see from \eqref{boundtra} that we can accomplish $e^{2\Phi} m^2 L_{int}^2 \ll 1$ by tuning the orientifold charge close to its bound.  
However, we must also make sure that the second square in \eqref{Wsmall} is small, which means
that $f g_s L_{int} \sim |\mathcal{W}_1^-| L_{int}$ is small. Manifolds for which $\mathcal{W}_1^-$ vanishes (and only
$\mathcal{W}_2^-$ is possibly non-zero) are called `nearly Calabi-Yau' (NCY) see e.g.~\cite{feng}; hence for the condition \eqref{Wsmall}
to be satisfied, the internal manifold must admit an SU(3)-structure which is sufficiently close to the NCY limit.

%%%%%%%%%%%%%%%%%%%%%%%%%%%%%%%%%%%%%%%%%%%%%%%%%%%%%%%%%%%

\subsection{Solutions on nilmanifolds and cosets}

By taking the internal six-dimensional space to be a
nilmanifold, it turns out that one can construct explicit examples of the
type of compactifications reviewed in section \ref{IIAsusycond}.
A systematic scan yields exactly two possibilities, namely the six-torus and the nilmanifold 4.7 of Table 4 of \cite{scan} (also known as the Iwasawa manifold), which (for some values of the parameters) turn out to be related by T-duality along two
directions.

For the torus let us define a left-invariant basis $\{e^i\}$ such that:
\eq{
\d e^i = 0, \qquad i=1,\dots, 6~.
}
We can just choose $e^i = \d y^i$, where $y^i$ are the internal coordinates.
The SU(3)-structure is given by
\begin{align}\label{jot}
J=e^{12}+e^{34}+e^{56} \, , \qquad \qquad
\Omega=(ie^1+e^2)\wedge(ie^3+e^4)\wedge(ie^5+e^6)~.
\end{align}
It readily follows that all torsion classes in \eqref{torsionclassesv} vanish in this case. However, there are non-vanishing $H$ and $F_{4}$
fields given by \eqref{ltsol}:
\eq{
\begin{split}
H  = \frac{2}{5} e^{\Phi} m \left(e^{246}-e^{136}-e^{145}-e^{235}\right) \, , \qquad
F_4  = \frac{3}{5} m \left( e^{1234} + e^{1256} + e^{3456} \right) \, .
\end{split}
}
{} From \eqref{boundtra} we find that there is an orientifold source with $\mu=e^{2\Phi} m^2$ and $w_3 = 0$, which corresponds to smeared orientifolds along $(1,3,5)$, $(2,4,5)$, $(2,3,6)$
and $(1,4,6)$.

For the  Iwasawa manifold the left-invariant basis is defined by:
\eq{
\d e^a=0,~~a=1,\dots, 4 \, ,  \qquad \d e^5=e^{13}-e^{24} \, , \qquad \d e^6=e^{14}+e^{23}~.
}
Up to basis transformations there
is a unique SU(3)-structure satisfying the supersymmetry conditions of section \ref{IIAsusycond}. It is given by
\eq{\spl{
J  = e^{12}+e^{34}+ \beta^2 e^{65} \, , \qquad \qquad
\Omega=\beta \, (ie^5-e^6)\wedge(ie^1+e^2)\wedge(ie^3+e^4)~,
\label{lolut}
}}
with metric $g=\text{diag} (1,1,1,1,\beta^2,\beta^2)$. Again we read off the  non-vanishing torsion classes from (\ref{torsionclassesv})
and  the fluxes from \eqref{ltsol} using (\ref{ltsolc}). We find from \eqref{boundtra} a non-zero net orientifold six-plane charge
$\mu\geq\frac{225}{16}|{\cal W}^-_1|^2$ ~. For the case $m=0$, for which this bound  is saturated, the above example can also be obtained by performing two T-dualities on the torus solution.

Another large class of IIA solutions of the type described in section \ref{IIAsusycond} is
given in \cite{klt}, which also incorporates certain solutions that were already known into
the single unifying framework of \emph{left-invariant} SU(3)-structures on coset spaces  $G/H$. 
Using the Maurer-Cartan equation and the commutation relations of the corresponding Lie algebras one finds for the exterior derivative of the globally defined one-forms
$
\d e^i = -\frac{1}{2}f^i_{jk} e^j \wedge e^k \, ,
$
where $f^i_{jk}$ are the structure \emph{constants} of the corresponding Lie algebras. 
The condition of left-invariance restricts the set of forms on a given coset. For example for $\frac{\text{G}_2}{\text{SU(3)}}$ the $G$-invariant two-forms and three-forms are spanned by
$\{e^{12}-e^{34}+e^{56}\}$ and $\{e^{245}+e^{135}+e^{146}-e^{236},-e^{235}-e^{246}+e^{145}-e^{136}\}$, respectively, and there are no invariant one-forms.
The most general solution is then given by
\eq{\spl{
J & = a (e^{12} - e^{34} + e^{56}) \, , \qquad \qquad a>0 \quad\text{(metric postivity)},\\
\Omega & = a^3 \left[ (e^{245}+e^{146}+e^{135}-e^{236}) + i (e^{145}-e^{246}-e^{235}-e^{136}) \right] \, ,
}}
where the overall scale is a free parameter. And again we read off the  solution from (\ref{torsionclassesv})-\eqref{ltsolc} and \eqref{boundtra}.

In \cite{klt} all six-dimensional cosets were scanned for solutions of type IIA AdS$_4$ compactifications with SU(3) structure. They found solutions on five different cosets, for each of which we will analyze the low energy effective theory.

\section{Low energy physics}\label{lowennil}

We will first explicitly perform a Kaluza-Klein reduction on the nilmanifolds and calculate the mass spectrum.  Next, we will use the effective supergravity approach and construct the K\"ahler potential and the superpotential. From there we can get the potential and compare the mass spectrum in both approaches. We find exact agreement. For the cosets we will only use the effective supergravity approach.

\subsection{Kaluza-Klein reduction}\label{seckkmain}

Let $x$ and $y$ be space-time and internal-manifold coordinates, respectively.
Moreover, let $\bg{\Phi}(x,y)$ be a `vacuum', i.e.\ a particular solution of
the equations of motion of ten-dimensional supergravity.
The Kaluza-Klein reduction (see e.g. \cite{duff} for a review) consists in expanding all ten-dimensional fields $\Phi(x,y)$ in `small' fluctuations $\delta \Phi(x,y)$ around the vacuum $\bg{\Phi}(x,y)$ keeping only  terms up to linear order in $\delta\Phi(x,y)$ in the equations of motion (corresponding to at most quadratic terms in the Lagrangian) and Fourier-expanding the fluctuations in the
internal space:
\al{\label{kkansatz}
\Phi(x,y)=\bg{\Phi}(x,y)+\delta\Phi(x,y)~ \, , \qquad \qquad 
\delta\Phi(x,y)=\sum_n\phi_n(x)\omega_n(y)
~,
}
where $\phi_n(x)$ are four-dimensional space-time fields,
and the
$\omega_n(y)$'s
form a basis of eigenforms of the Laplacian operator $\Delta=\d \d^\dagger+\d^\dagger \d$ in the six-dimensional space $\mathcal{M}$ (the internal
part of the vacuum solution). In the following we will truncate
all the higher Kaluza-Klein modes in the harmonic expansion and keep
only those $\omega_n(y)$'s that are left-invariant
on $\mathcal{M}_6$. The resulting modes are not in general harmonic, but correspond to
eigenvectors of the Laplacian whose eigenvalues are of order of the geometric fluxes.
Plugging the ansatz (\ref{kkansatz}) into the ten-dimensional equations of motion and keeping at most linear-order
terms in the fluctuations, one can read off the masses of the space-time fields, i.e.\ the `spectrum'.
In the present case, this is accomplished by comparing with the equations of motion
for non-interacting
fields propagating in AdS$_4$. For scalars one gets \cite{duff}
\al{
\label{scalar}\Delta\phi=- \left(M^2+\frac{2}{3}\Lambda_{\text{AdS}}\right)\phi \equiv -\tilde{M}^2 \phi~.
}
The Breitenlohner-Freedman bound \cite{bf} is given by
$
\tilde{M}^2 \geq -\frac{9|W|^2}{4} \, .
$
We will take $\tilde{M}=0$ as the definition of an unstabilized modulus since
from \eqref{scalar} we see that then, if it were not for the boundary conditions of
AdS$_4$, a constant shift of $\phi$ would be a solution to the equations of motion.

\subsection{Effective supergravity}

The scalar potential is given in terms of the superpotential and K\"ahler potential via
\eq{\label{V}
V(\phi,\bar{\phi}) = M_P^{-2} e^{\mathcal{K}} \left( \mathcal{K}^{i\bar{\jmath}} D_i \mathcal{W}_{\E} D_{\bar{\jmath}} \mathcal{W}^*_{\E} - 3 |\mathcal{W}_{\E}|^2 \right) \, .
}
The superpotential and K\"{a}hler potential of the effective $\mathcal{N}=1$ supergravity
have been derived in various ways in \cite{granasup,grimmsup,effective} (based on earlier work of \cite{GVW,gl}).
The superpotential in the Einstein frame $\mathcal{W}_{\E}$ reads
for the IIA SU(3) case with pure spinors
\eq{\label{WSU3}
\mathcal{W}_{\E}  = \frac{-i e^{-i \theta}}{4 \kappa_{10}^2} \int_M \langle e^{i(J-i\delta B)}, \bg{F} - i~\d_{\bg{H}} \left( e^{\delta B} e^{-\Phi} \Im \Omega +i\delta C_3 \right) \rangle \, ,
}
where $\langle \cdot, \cdot \rangle$ indicates the Mukai pairing 
$
\label{mukai}
\langle \phi_1, \phi_2 \rangle = \phi_1 \wedge \alpha(\phi_2)|_{\text{top}} \, 
$
and the operator $\alpha$ acts by inverting the order of indices on forms.
The K\"ahler potential is given by
\eq{\label{KahlerJOmega}
\mathcal{K}  = - \ln \int_M \, \frac{4}{3} J^3 - 2 \ln \int_M \, 2 \, e^{-\Phi}\Im \Omega \wedge e^{-\Phi} \Re \Omega + 3 \ln(8 \kappa_{10}^2 M_P^2) \, ,
}
where $e^{-\Phi }\Re \Omega$ should be seen as a function of $e^{-\Phi}\Im \Omega$ \cite{effective}.
On the fluctuations we must impose the orientifold projections. By expanding in a suitable basis of even and odd expansion forms (which have to be identified separately for each case), we find that the fluctuations organize naturally in complex scalars:
\begin{eqnarray}\label{expansionSU3}
J - i \delta B &=& (k^i - i b^i)Y^{(2-)}_i = t^i Y^{(2-)}_i \, , \nn \\
 e^{-\Phi} \Im \Omega + i \delta C_3 &=& (u^i +i c^i) e^{-\hat{\Phi}} Y^{(3+)}_i = z^i e^{-\hat{\Phi}} Y^{(3+)}_i \, ,
\end{eqnarray}
where we took out the background $e^{-\hat{\Phi}}$ from the definition of $z^i$ for further
convenience.

\subsection{Effective theory of nilmanifolds and cosets}

\label{torusKKresult}

By direct computation of the Kaluza-Klein reduction on the nilmanifolds we obtain for the torus exactly the same mass spectra as for the Iwasawa.\footnote{The interested reader may consult \cite{adseff} for more details on the derivation and on the exact mass eigenvalues and eigenvectors.} This is of course the expected result, since the two solutions are related by T-duality. All three axions stay massless as expected. The complex structure moduli are tachyonic but stable, because they are still above the Breitenlohner-Freedman bound.  
Scalars that are in the same supermultiplet have different masses due to a subtlety of the supersymmetry algebra of AdS$_4$, which no longer  allows a definition for the mass as an invariant Casimir operator.
For these models, we can decouple the tower of Kaluza-Klein masses when we take $m^2 (e^{2\Phi} L_{int}^2)\ll1$ for the torus and $\beta \ll 1$ for the Iwasawa.

Using the effective supergravity approach we obtain the same results. After choosing the odd two- and even three-forms in (\ref{expansionSU3}) it is straightforward to compute  the superpotential (\ref{WSU3}) and K\"{a}hler potential \eqref{KahlerJOmega} for the torus and the Iwasawa. Actually the K\"ahler potential is the same in both cases while we find for the superpotentials
\eq{
\mathcal{W}_{\E, \text{Iwasawa}} = -i t^1 \mathcal{W}_{\E, \text{Torus}}(t^1 \rightarrow \frac{1}{t^1}) \, ,
}
which is expected from T-duality. Plugging the results into \eqref{V} on can calculate the masses for the scalar fields. We find exactly the same result as for the KK-reduction. The agreement of the two aproaches provides a consistency check on the ability of the effective supergravity approach to handle geometric fluxes. For the coset spaces we will only use the latter one to compute the masses of the scalar fields.

For each coset we find the expansion forms in \eqref{expansionSU3} by imposing the orientifold involutions on the set of left invariant forms. Then we compute the potential \eqref{V} in terms of the superpotential \eqref{WSU3} and the K\"ahler potential \eqref{KahlerJOmega}. The following table lists the coset spaces found in \cite {klt}, indicating in each case
the number of light real scalar fields, the number of them that stay massless and whether it is possible to decouple the
tower of Kaluza-Klein modes in the AdS vacuum.
\begin{center}
\begin{tabular}{|c|c|c|c|c|c|}
\hline
& \rule[1.2em]{0pt}{0pt} $\frac{\text{G}_2}{\text{SU(3)}}$& $\frac{\text{Sp(2)}}{\text{S}(\text{U(2)}\times \text{U(1)})}$ & $\frac{\text{SU(3)}}{\text{U(1)}\times \text{U(1)}}$ & SU(2)$\times$SU(2) & $\frac{\text{SU(3)}\times \text{U(1)}}{\text{SU(2)}}$\\
\hline
Light fields & 4 & 6 & 8 & 14 & 8 \\
Unstabilized & 0 & 0 & 0 & 1 & 0 \\
Decouple KK & no & yes & yes & yes & no\\
\hline
\end{tabular}
\label{cosettable}
\end{center}
All moduli are stabilized in each model except for SU(2)$\times$SU(2). However, it turns out to be rather hard to
decouple the tower of Kaluza-Klein modes and in only three models there is a limit where this happens. However, an additional uplift term may also help to decouple the Kaluza-Klein modes.

\section*{Acknowledgments}
It is a pleasure to thank C. Caviezel, P. Koerber, D. L\"ust, D. Tsimpis and M. Zagermann for the collaboration on \cite{adseff}. Further, I want to thank the organizers of the 4th RTN workshop for giving me the opportunity to present this work.


\begin{thebibliography}{[10]}

\bibitem{adseff}% article
 C.~Caviezel,  P.~Koerber,  S.~K\"ors,
  D.~L\"ust,  D.~Tsimpis,  and  M.~Zagermann, "The effective theory of type IIA AdS4 compactifications on nilmanifolds and
  cosets"
  [arXiv:0806.3458].


\bibitem{kt}% article
 P.~Koerber and  D.~Tsimpis, "Supersymmetric sources,
  integrability and generalized- structure compactifications",
 JHEP \textbf{08}, 082 (2007)
  [arXiv:0706.1244].


\bibitem{lt}% article
 D.~L\"ust and  D.~Tsimpis, "Supersymmetric AdS(4)
  compactifications of IIA supergravity",
 JHEP \textbf{02}, 027 (2005)
 [arXiv:hep-th/0412250].


\bibitem{gencal1}% article
 P.~Koerber, "Stable D-branes, calibrations and generalized
  Calabi-Yau geometry",
 JHEP \textbf{08}, 099 (2005)
 [arXiv:hep-th/0506154].


\bibitem{gencal2}% article
 L.~Martucci and  P.~Smyth, "Supersymmetric D-branes
  and calibrations on general N = 1 backgrounds",
 JHEP \textbf{11}, 048 (2005)
 [arXiv:hep-th/0507099].


\bibitem{granaN1}% article
 M.~Gra\~na,  R.~Minasian,  M.~Petrini,  and
  A.~Tomasiello, "Generalized structures of N=1 vacua",
 JHEP \textbf{11}, 020 (2005)
 [arXiv:hep-th/0505212].


\bibitem{gualtieri}% article
 M.~Gualtieri~, "Generalized complex geometry" 
 Oxford University DPhil thesis (2003)
 [arXiv:math.DG/0401221].


\bibitem{roma}% article
 L.\,J. Romans, "Massive N=2a Supergravity in
  Ten-Dimensions",
 Phys. Lett. \textbf{B169}, 374 (1986).


\bibitem{democratic}% article
 E.~Bergshoeff,  R.~Kallosh,  T.~Ort\'{\i}n,
  D.~Roest,  and  A.~Van~Proeyen, "New formulations of
  D = 10 supersymmetry and D8 - O8 domain walls",
 Class. Quant. Grav. \textbf{18}, 3359--3382 (2001)
 [arXiv:hep-th/0103233].


\bibitem{tomtwistor}% article
 A.~Tomasiello, "New string vacua from twistor spaces",
 Phys. Rev. \textbf{D78}, 046007 (2008)
 [arXiv:0712.1396].


\bibitem{feng}% article
 F.~Xu, "SU(3)-structures and special lagrangian geometries",
 [arXiv:math.DG/0610532].


\bibitem{scan}% article
 M.~Gra\~na,  R.~Minasian,  M.~Petrini,  and
  A.~Tomasiello, "A scan for new N=1 vacua on twisted tori",
 JHEP \textbf{05}, 031 (2007)
 [arXiv:hep-th/0609124].


\bibitem{klt}% article
 P.~Koerber,  D.~L\"ust,  and  D.~Tsimpis, "Type IIA AdS4 compactifications on cosets, interpolations and domain
  walls",
 JHEP \textbf{07}, 017 (2008)
  [arXiv:0804.0614].


\bibitem{duff}% article
 M.\,J. Duff,  B.\,E.\,W. Nilsson,  and  C.\,N.
  Pope, "Kaluza-Klein Supergravity",
 Phys. Rept. \textbf{130}, 1--142 (1986).


\bibitem{bf}% article
 P.~Breitenlohner and  D.\,Z. Freedman, "Stability in
  Gauged Extended Supergravity",
 Ann. Phys. \textbf{144}, 249 (1982).


\bibitem{granasup}% article
 M.~Gra\~na,  J.~Louis,  and  D.~Waldram, "SU(3) x SU(3) compactification and mirror duals of magnetic fluxes",
 JHEP \textbf{04}, 101 (2007)
 [arXiv:hep-th/0612237].


\bibitem{grimmsup}% article
 I.~Benmachiche and  T.\,W. Grimm, "Generalized N = 1
  orientifold compactifications and the Hitchin functionals",
 Nucl. Phys. \textbf{B748}, 200--252 (2006)
 [arXiv:hep-th/0602241].


\bibitem{effective}% article
 P.~Koerber and  L.~Martucci, "From ten to four and
  back again: how to generalize the geometry",
 JHEP \textbf{08}, 059 (2007)
 [arXiv:0707.1038].


\bibitem{GVW}% article
 S.~Gukov,  C.~Vafa,  and  E.~Witten, "CFT's
  from Calabi-Yau four-folds",
 Nucl. Phys. \textbf{B584}, 69--108 (2000)
 [arXiv:hep-th/990670].


\bibitem{gl}% article
 T.\,W. Grimm and  J.~Louis, "The effective action of
  type IIA Calabi-Yau orientifolds",
 Nucl. Phys. \textbf{B718}, 153--202 (2005)
 [arXiv:hep-th/0412277].


\end{thebibliography}
\end{document}